\documentclass[nofootinbib]{revtex4-2}

\usepackage[english]{babel}

\usepackage[a4paper,top=2cm,bottom=2cm,left=3cm,right=3cm,marginparwidth=1.75cm]{geometry}

\usepackage{amsmath}
\usepackage{amssymb}
\usepackage{graphicx}
\usepackage[colorlinks=true, allcolors=blue]{hyperref}
\usepackage{color}
\usepackage{xcolor}
\usepackage{comment}
\usepackage{orcidlink}

\begin{document}

\title{IR side of bounds on Theories with Spontaneously Broken Lorentz Symmetry}

\author{Francesco Serra}
\email[]{fserra2@jh.edu}
\affiliation{Department of Physics and Astronomy, Johns Hopkins University, Baltimore, MD 21218, USA}
	
\author{Leonardo G. Trombetta}
\email[]{trombetta@fzu.cz}
\affiliation{CEICO, Institute of Physics of the Czech Academy of Sciences, Na Slovance 1999/2, 182 00, Prague 8, Czechia,}


\begin{abstract}
In nature, some UV features of dynamics are reflected in IR quantities.
In fully relativistic theories, this connection can be probed through the analyticity properties of scattering amplitudes, allowing one to understand which IR theories respect the UV assumptions of quantum field theory. The ensuing analyticity bounds can usually be rephrased as the absence of faster-than-light propagation for low-energy excitations.
While it is interesting to understand these relations and their IR characterization for theories that have less idealized properties, it is also more difficult to derive analyticity bounds in these cases. For theories that spontaneously break Lorentz symmetry, recent progress was made by considering correlators of conserved currents and their analyticity properties. In this work, we focus on such theories and work to close the gap from the IR side, finding a natural way to express the known analyticity bounds purely in terms of low-energy kinematical quantities. Our analysis shows that the bounds require gapped excitations to have a slower speed than the gapless ones, at least for momenta that are low with respect to the mass gap. These results suggest a way to interpret the UV/IR connection in more complex theories.
\end{abstract}

\maketitle
\flushbottom

\section{Introduction}

	Separation of scales in nature allows one to describe low-energy (IR) dynamics without knowledge of high-energy (UV) degrees of freedom. At the same time, low-energy interactions encode some of the properties that characterize the high-energy dynamics. 

	This connection between UV and IR becomes remarkably transparent for theories that in the UV are compatible with unitarity, Lorentz invariance, locality and microcausality -- the property that operators commute when space-like separated \cite{Eden1971}. These assumptions about the UV theory entail polynomial boundedness and canonical analyticity properties of S-matrix elements as functions of the Mandelstam invariants \cite{Bros1964,Martin:1969,Streater:1989}, leading to bounds on the coefficients of low-energy interactions, see e.g. \cite{Pham:1985,Adams:2006sv,deRham:2017avq,Caron-Huot:2020cmc,Bellazzini:2020cot,Tolley2021,Sinha:2020win,Bellazzini2022,Arkani-Hamed2021a,Caron-Huot2022}.

	In light of this UV/IR connection, it is interesting to explore whether these boundaries in the space of low-energy Wilson coefficients can be characterized in a synthetic way in terms of IR dynamical features.
	Such an IR characterization has been explored e.g. in \cite{Adams:2006sv,Camanho:2014apa,deRham:2021bll,Serra2022,CarrilloGonzalez2022,CarrilloGonzalez2023,Serra:2023nrn}, finding that certain effective field theories (EFTs) that do not respect the analyticity bounds display faster-than-light propagation for excitations around non-trivial backgrounds.
    These results exposed that the speed of propagation of low-energy excitations captures, at least in some cases, the UV properties that lead to analyticity bounds. In particular, in various examples of Lorentz invariant theories, one can argue (and sometimes prove) that analyticity bounds forbid classical time advances at least on a wide class of backgrounds; see e.g. \cite{Arkani-Hamed2021a,Caron-Huot2022,Bellazzini:2022wzv}. 
    
	This may be phrased in terms of the relation between microcausality and the kinematics of low-energy excitations. Indeed, the limiting value of the phase velocity at high frequency, also known as the front/signal velocity, should be insensitive to any fixed wavelength background and should therefore fall into the light-cone as dictated by microcausality \cite{Brillouin:1960tos,Diener:1996mj}. Similarly, in the regime where signal propagation can effectively be described as classical trajectories/rays, the group velocity encodes the support of Green's functions \cite{Diener:1996mj}.
    Therefore, for cases in which group and/or phase velocity do not decrease in the UV, their values at low frequency should be lower than the speed determined by microcausality\footnote{See e.g., \cite{Diener:1996mj,Greene:2022uyf} for interesting exceptions in this regard.}.
    
    Given these results, it is valuable to understand how the same connection between UV and IR becomes manifest in more general contexts, where Lorentz invariance, locality, or other assumptions may be to some extent relaxed, in either or both the UV and the IR.     
    For instance, EFTs describing physical systems like condensed matter or cosmology may carry information about the underlying UV theory, even if the low-energy interactions nonlinearly realize Lorentz invariance. 
    As a matter of fact, analyticity bounds are rooted in the study of propagation of light in a material, see e.g. \cite{Keldysh1989TheDF,Creminelli:2024lhd} and have recently been fruitfully extended to more generic theories that spontaneously break Lorentz invariance \cite{Creminelli:2022onn}. 
    Similarly, less conservative assumptions about the UV may still entail rich analyticity structures and map out boundaries in the space of Wilson coefficients that might be measured in the future, see e.g. \cite{Haring:2022cyf,Buoninfante:2023dyd, Buoninfante:2024ibt} for recent efforts in this direction.
    In this regard, it would be very interesting to understand whether analyticity bounds and the corresponding UV assumptions have a universal translation in terms of IR dynamical features.
    
    In this work, as a first step toward understanding the IR characterization of analyticity bounds beyond the most conservative assumptions, we focus on the case of an EFT that spontaneously breaks Lorentz symmetry. 
    In this context, the analyticity of correlators of conserved currents has been exploited to obtain bounds on the coefficients of low-energy interactions \cite{Creminelli:2022onn}.
    We examine these results by studying how the propagation of low-energy excitations in the EFT is affected by interactions, expressing the known bounds on the Wilson coefficients in terms of the excitations' kinematics.
    In this way, we provide an IR interpretation of the analyticity bounds derived in \cite{Creminelli:2022onn}, showing that a specific notion of velocity for low-energy excitations sharply captures the information contained in the analyticity bounds. 
    In fact, our result is that the acoustic vector, defined in terms of the effective geometry that describes propagation at low energy and the momentum (wave-number) co-vector, must lay inside a fixed, momentum-independent, sound-cone. This remarkably simple condition, characterizes synthetically and in a momentum-independent way the analyticity bounds in terms of the IR kinematics.
    
    Specifically, we consider the EFT of a conformal superfluid coupled to a gauge field in $D=3$ space-time dimensions. These systems are characterized by a non-linear realization of time translations due to a finite charge density phase of a scalar field charged under $U(1)$. This phase breaks the conformal symmetry to spatial rotations and translations plus the non-linearly realized time-translation. The IR dynamics of these systems has been studied extensively in the context of large charge expansion, see e.g. \cite{Hellerman:2015nra,Monin:2016jmo,Cuomo:2020rgt,Cuomo:2021qws}. We stress that this choice of EFT solely follows the need for a tractable example for which analyticity bounds have been derived, and is to be considered as a proxy for more complicated {situations} such as theories relevant for modeling cosmology. In the same vein, our studies here are intended as a first step towards a more complete IR {characterization} of analyticity bounds.

    In these systems, the existence and analyticity properties of the S-matrix become nontrivial \cite{Hui:2023pxc,Creminelli:2023kze}, requiring one to consider alternative approaches in order to derive analyticity bounds. In \cite{Creminelli:2022onn} this was done by studying correlators of the conserved $U(1)$ currents, due to their convenient UV behavior and analytic properties. In practice, these correlators were computed by adding a background, non-dynamical $U(1)$ gauge field coupled to the superfluid.
    While the $U(1)$ symmetry can be easily gauged, it is not simple to study the dynamics of the gauge field and to prospect which UV completions would give rise to the gauged EFT in the broken phase. 
    
    In our work, we follow an approach similar in spirit to what was done in \cite{Creminelli:2022onn} and introduce a $U(1)$ gauge field coupled to the conformal superfluid, with vanishing background field-strength. 
    Similarly to \cite{Creminelli:2022onn}, the gauge field is considered as an auxiliary system to extract key information about the conformal superfluid. While \cite{Creminelli:2022onn} considers the gauge field as fixed background, we keep track of its kinematics. In fact, it is natural to regard the background field as an approximation of a dynamical field whose excitations are being neglected. 
    While it might be subtle to interpret this IR system as a physical system that is UV completed by a conformal field theory, we imagine that a background charge density makes the configuration with vanishing background field strength stable. As a matter of fact, the quadratic part of the effective action for the $U(1)$-gauged conformal superfluid allows to consistently describe the excitations of both the scalar and gauge fields, even when one writes this action imagining a non-dynamical, background gauge field. 
    In this action, a kinetic mixing converts scalar modes into modes of the gauge field and viceversa.
    As we discuss, this kinetic mixing allows one to recover the information contained in correlators of conserved currents by examining the propagation of the low-energy modes.
    
    A universal feature of the EFTs for the broken phase is that gapless excitations at zero momentum have soundspeed $c_s^2=1/2$ in $D=3$. This means that in any UV complete model that has Lorentz symmetry in the UV, the phase velocity of a gapless mode will change from $1/2$ to $1$ as one explores excitations with higher and higher momenta (eventually outside of the regime of validity of the EFT for the broken phase). If the low-energy theory has a mass gap, the phase velocity cannot be used to characterize the system at momenta below the mass gap, as it is divergent at zero momentum. In these cases, one can use a different notion of velocity associated with wave packets \cite{Boillat:1969slt,Boillat:1970gw,Landau,Sawicki:2024ryt}, the group velocity. This coincides with the phase velocity for several gapless systems, but keeps being well-defined for gapped excitations at zero momentum. Here we find that neither of these notions of velocity can be used to express the known analyticity bounds in a sharp, momentum-independent way. This difficulty prompts us to instead study directly the propagation through the effective acoustic metric seen by IR excitations. In this language, an acoustic vector can be naturally defined, leading to a notion of velocity which is closely related to both the phase and the group velocity, but that differs from these two in general. (In fact, as we will show, this new velocity matches the phase velocity for gapless excitations, while it coincides with the group velocity in other cases.) From this point of view, we find that the analyticity bounds are encoded in a momentum-independent condition on the acoustic vector of IR excitations and its associated velocity. Physically, this condition requires gapped excitations to have a lower velocity than the gapless ones, for momenta below the mass gap. We apply this reasoning to the gauged conformal superfluid, since the kinetic mixing between scalar and gauge-field excitations leads to a mass gap for all the physical modes. In practice, we ask the velocity associated to the acoustic vector to be at most $1/2$ for momenta lower than the mass gap. In this way, the excursion between IR and UV values of the velocity is entirely captured by the (effectively) gapless, high momentum regime of the theory where the velocity defined by the acoustic vector asymptotes to the phase velocity, as one might expect in analogy with the gapless cases. This IR requirement, as we show, leads to the same bounds found in \cite{Creminelli:2022onn} studying correlators of conserved currents.
    
    This result makes it clear that the IR meaning of the analyticity bounds is not the absence of paradoxical causality violations, as previously thought \cite{Adams:2006sv}. Indeed, propagation outside the speed $1/2$ soundcone is not in conflict with causality, while it is forbidden by the analyticity bounds. This is consistent with the finding that no causal paradoxes associated with superluminality in the IR can be predicted in the EFT, as discussed in \cite{Babichev:2007dw,Burrage:2011cr,Kaplan:2024qtf}.
    The work is organized as follows. In Section 2 we describe the model of gauged conformal superfluid and its propagating modes at low frequencies. In Section 3 we discuss the IR kinematics for gapless and gapped modes using the acoustic metric description, introducing and motivating various natural notions of velocity. In Section 4, we consider the analyticity bounds of \cite{Creminelli:2022onn} on the EFT of conformal superfluids, expressing them as a momentum-independent condition on the acoustic vector of low-energy excitations. We close in Section 5 with a brief discussion. Throughout this paper we use the $(-,+,\dots,+)$ signature convention. We refer to vectors $v = v^\mu \partial_\mu$ and co-vectors $w = w_\mu dx^\mu$, respectively, in terms of their contravariant $v^\mu$ and covariant $w_\mu$ components, with $\{ \partial_\mu \}$ and $\{ dx^\mu \}$ being the corresponding coordinate basis of the tangent and co-tangent spaces.

\section{Conformal superfluid coupled to a gauge field}\label{sec:superf}

We start by considering an effective field theory for a conformal superfluid in $D=3$ coupled to a gauge field, as discussed in Ref.~\cite{Creminelli:2022onn},
\begin{eqnarray} \label{lagrangian}
\mathcal{L} &=& \frac{c_1}{6} |\nabla \chi|^3 - 2 c_2 \frac{(\partial |\nabla \chi|)^2}{|\nabla \chi|} + c_3 \left( 2 \frac{(\nabla^\mu \chi \partial_\mu |\nabla \chi|)^2}{|\nabla \chi|^3} + \partial_\mu \left( \frac{\nabla^\mu \chi \nabla^\nu \chi}{|\nabla \chi|^2} \right) \partial_\nu |\nabla \chi| \right) \notag \\ 
&& - \frac{b}{4} \frac{F_{\mu\nu} F^{\mu\nu}}{|\nabla \chi|} + \frac{d}{2} \frac{F^{\mu}{}_{i} F^{\nu i}}{|\nabla \chi|^3} \nabla_\mu \chi \nabla_\nu \chi \, ,
\end{eqnarray}
where
\begin{eqnarray}
    \nabla_\mu \chi &\equiv& \partial_\mu \chi - A_\mu \, , \\
    |v| &\equiv& \sqrt{- v_\mu v^\mu} \,.
\end{eqnarray}
As we outline in Sec.~\ref{sec:UV-compl}, such an EFT arises e.g. when integrating out the radial mode of a $U(1)$ charged scalar in a conformal field theory. If the gauge field was a dynamical degree of freedom in the UV, then one would expect a purely kinetic term for the gauge field to appear in Eq.~\eqref{lagrangian}, for instance derived using the symmetry arguments in \cite{Cuomo:2021qws}. To keep clear the analogy with the work in \cite{Creminelli:2022onn}, we will neglect this term and use the above EFT only as an auxiliary IR system to extract information about the conformal superfluid. As we show below, the interactions between scalar and gauge field already endow the latter with a healthy kinetic term in the broken phase. This theory only makes sense on top of a time-dependent background for the scalar field
\begin{equation} \label{LB-bg}
    \chi(x) = \mu t + \pi(x) \,,
\end{equation}
which is responsible for the spontaneous breaking of Lorentz invariance below the scale $\mu$, which we assume as $\mu > 0$ without loss of generality, and which acts as the cutoff of the EFT for the broken phase. This theory enjoys a gauge symmetry
\begin{eqnarray}
    \pi \to \pi + \Lambda \qquad ; \qquad A_\mu \to A_\mu + \partial_\mu \Lambda \, .
\end{eqnarray}

In order to study the dynamics of the low-energy degrees of freedom in the broken phase we turn to the quadratic Lagrangian for both $\pi$ and $A_\mu$,
\begin{eqnarray} \label{lagrangian2}
    \mathcal{L}_2 &=& \frac{c_1 \mu^3}{6} + \frac{\mu c_1}{2} \left[ \left( \dot{\pi} + A^0 \right)^2 - \frac{1}{2} \left( \partial_i \pi - A_i \right)^2 + \mu \left( \dot{\pi} + A^0 \right) \right] + \frac{2c_2}{\mu} \left[ -\pi \square \ddot\pi + 2 A^0 \square \dot\pi + A^0 \square A^0 \right] \notag \\
    &&+ \frac{2c_3}{\mu} \left[ -\pi \square \ddot\pi + 2 A^0 \square_{c_s} \dot\pi - A^i \partial_i \ddot\pi + (\dot{A}^0)^2 + \dot{A}^0 \partial_i A^i \right] \notag \\
    &&+ \frac{(b+d)}{2\mu} \left[ (\partial_i A^0)^2 + (\partial_0 A_i)^2 + 2 \dot{A}^0 (\partial_i A_i) \right] - \frac{b}{4\mu} (\partial_i A_j - \partial_j A_i)^2 \, ,
\end{eqnarray}
with $i=1,2$ running over the spatial directions. This Lagrangian provides both kinetic terms for $\pi$ and for $A_\mu$, which are free of ghost, gradient and tachyonic instabilities for $c_1 > 0$, $b > 0$ and $b + d > 0$. Therefore, this quadratic Lagrangian and the related Hamiltonian, consistently describe the gauge and scalar fields as healthy, dynamical variables. Importantly, there is also kinetic mixing between $\pi$ and $A_\mu$ which implies that if one excites either of the two fields, inevitably the other will get excited as well. Due to this, we should study the eigenmodes of propagation.

As discussed in the Introduction, we regard the gauge field as part of an IR auxiliary system we use to study the conformal superfluid. We imagine that the IR system describing the broken phase with vanishing electric potential is stable due to a suitable homogeneous background charge. This background charge makes it possible for the scalar field to get a non-zero charge density while satisfying the Gauss law with zero background electric field. This phase will be stable if one chooses appropriate self interactions for the scalar.

The coefficients $c_2$ and $c_3$ of the Higher-Dimensional (HD) operators have been shown to be restricted by analyiticity to satisfy \cite{Creminelli:2022onn}
\begin{eqnarray} \label{bound-xi}
    \frac{c_2}{b+d} \left( 1 - \xi^2 \right) - \frac{c_3}{b+d} \geq - \frac{\left( 1 - \frac{\xi^2}{2} \right)^2}{\xi^2} \, ,
\end{eqnarray}
where $\xi \in [0,1]$. We are interested in expressing this bound in terms of IR kinematic quantities, for which we proceed to study the propagating degrees of freedom.

\subsection{Scalar gapless mode} \label{sec:scalar-mode}

It is instructive to first consider the decoupling limit when the gauge field is in fact nondynamical (formally $b+d \to \infty$). In this case, the dispersion relation for the canonical scalar $\pi_c = \sqrt{\mu c_1} \pi$ is obtained as
\begin{eqnarray} \label{scalar-mode-cond}
    G^{-1}_{\pi\pi}(\omega, {\bf p}) &=& \left( \omega^2 - \frac{{\bf p}^2}{2} \right) + \frac{4 (c_2 + c_3)}{\mu^2 c_1} \left( \omega^2 - {\bf p}^2 \right) \omega^2 = 0 \, .
\end{eqnarray}
When solved perturbatively in $c_2$ and $c_3$ at linear order, this gives
\begin{eqnarray} \label{massless-decoupled-mode}
    \omega_\pi^2({\bf p}) = \frac{{\bf p}^2}{2} + \frac{4(c_2+c_3)}{\mu^2 c_1} \left( \frac{{\bf p}^2}{2} \right)^2 \, .
\end{eqnarray}
As expected, due to the conformal symmetry in $D=3$ and the breaking of Lorentz invariance below $\mu$, the speed of propagation at zero momentum is $c_s^2 = 1/2$. In general we find a momentum dependent phase velocity\footnote{In this case the usual definition holds since the background is spatially isotropic. A more general definition is provided in Ref.~\cite{Sawicki:2024ryt}.} {$\omega_\pi^2({\bf p})/{\bf p}^2$}, which either increases or decreases with respect to $1/2$ depending on the sign of $c_2 + c_3$. 
At this point, we only notice that the phase velocity will increase monotonically for increasing (small) momenta only if $c_2 + c_3 > 0$. 
However, there is no further bound we can impose at this stage on the size of these coefficients, other than the requirement of validity of the perturbative treatment within the EFT, which is rather uninformative. Importantly, the coefficients $c_2$ and $c_3$ only appear here in the specific combination $c_2 + c_3$, an indication that the information contained in the bound \eqref{bound-xi} cannot be captured by the scalar mode alone. This prompts us to consider the dynamical nature of the gauge field in the search for an accurate representation of the analyticity bound.

\subsection{Dynamical gauge field}

 As discussed before, in general, for finite $b$ and $d$, the quadratic Lagrangian \eqref{lagrangian2} describes not only a propagating scalar, but also a propagating gauge field. In fact, these are kinetically mixed, and therefore, the eigenmodes of propagation are a combination of both. Interestingly, this endows the would-be scalar mode with a gap. 

We proceed to find the propagating degrees of freedom. The second order Lagrangian can be written, in Fourier space, as 
\begin{eqnarray}\label{eq:eomZ}
    \mathcal{L}_2 = \frac{1}{2} \Phi_I^\dag(-\omega, -{\bf p}) \, \mathcal{K}_{IJ}(\omega, {\bf p}) \Phi_J(\omega, {\bf p}) \, ,
\end{eqnarray}
where $\Phi_I = \left( \bar{\pi}, \bar{A}_\nu \right)$, and $\bar{\pi} = \sqrt{\mu} \, \pi$ and $\bar{A}_{\nu} = \frac{1}{\sqrt{\mu}} A_\nu$ are the dimension-$1/2$ fields, with the indices $I,J = \{ \chi, 0, 1, 2 \}$ running over the fields and their components. The kinetic $4\times4$ matrix $\mathcal{K}_{IJ}(\omega, {\bf p})$ is given by
\begin{eqnarray} \label{kinetic-matrix}
    \mathcal{K}_{IJ}(\omega, {\bf p}) = \left( 
    \begin{array}{c | c c }
        c_1 G^{-1}_{\pi\pi} & i \hat{J}_0 & i \hat{J}_j \\
        \hline
        -i \hat{J}_0 & u & v \, p_j \\
        -i \hat{J}_i & v \, p_i & V \delta_{ij} + b \, p_i p_j \\
    \end{array}
    \right)
    \, ,
\end{eqnarray}
which is organized into $1\times1$, $1\times3$, $3\times1$ and $3\times3$ blocks for clarity. The lower-case indices $i,j = \{1,2\}$ run instead over the spatial coordinates. The various quantities introduced here read as
\begin{eqnarray} 
    \hat{J}_0 &=& \left[ \mu^2 c_1 + 4 c_2 \left( \omega^2 - {\bf p}^2 \right) + 4 c_3 \left( \omega^2 - \frac{{\bf p}^2}{2} \right) \right] \frac{\omega}{\mu} \, , \label{hatJ0} \\
    \hat{J}_i &=& - \left( \frac{\mu^2 c_1}{2} + 2 c_3 \, \omega^2 \right) \frac{p_i}{\mu} \, , \label{hatJi} \\
    u &=& \mu^2 c_1 + 4 c_2 \left( \omega^2 - {\bf p}^2 \right) + 4 c_3 \, \omega^2 + (b + d) {\bf p}^2 \, , \\
    v &=& - \left( 2 c_3 + (b + d) \right) \omega \, , \\
    V &=& - \frac{\mu^2 c_1}{2} + (b + d) \omega^2 - b \, {\bf p}^2 \, .
\end{eqnarray}

The dispersion relations $\omega^2({\bf p})$ for the propagating degrees of freedom are obtained from the condition $\det( \mathcal{K} ) = 0$. To make sense of this procedure, we fix a ``Coulomb''-type gauge $\partial_i A^i = 0$. Other choices of gauge give the same result. We find the physical degrees of freedom to be determined by the following equations
\begin{eqnarray}
    0 &=& V \, , \label{transverse-mode-eq} \\     
    0 &=& u \left( V + b {\bf p}^2 \right) - v^2 {\bf p}^2 \, . \label{pi-A0-mode-eq}     
\end{eqnarray}
The first equation describes the transverse mode, with the dispersion relation
\begin{eqnarray}
    \omega_\perp^2({\bf p}) &=& m^2 + \frac{b}{(b + d)} {\bf p}^2 \,,
\end{eqnarray}
where
\begin{eqnarray}
    m^2 = \frac{\mu^2 c_1}{2(b+d)} \, .
\end{eqnarray}
Since there are no contributions from the HD operators $c_2$ and $c_3$ in $\omega_\perp^2({\bf p})$, we will not examine this further. The second one is a mix of the scalar and $A^0$ modes (after removing the longitudinal mode by the gauge fixing condition). In this second dispersion relation the size of $c_2$ and $c_3$ controls contributions with higher powers in $\omega$ and/or ${\bf p}$. Due to this, Eq.~\eqref{pi-A0-mode-eq} contains in fact both low-energy and high-energy modes. Focusing on the low-energy mode, we can solve perturbatively at linear order in $c_{2,3}$ within the regime of validity of the EFT. This leads to
\begin{eqnarray}
    \omega_{-}^2({\bf p}) &=& m^2 + \frac{{\bf p}^2}{2} - \frac{4(c_2+c_3)}{\mu^2 c_1} \left( - \frac{{\bf p}^2}{2} + m^2 \right) \left( \frac{{\bf p}^2}{2} + m^2 \right) \notag \\ 
    &&+ \frac{\mu}{2(b+d)} \left[ \frac{4c_2}{\mu} \left( - \frac{{\bf p}^2}{2} + m^2 \right) + \frac{4c_3}{\mu} \left( \frac{{\bf p}^2}{2} + m^2 \right) \right] \,.
\end{eqnarray}
Notice that when $c_1 \ll (b+d)$, the gauge effectively decouples from the scalar and also $m^2 \ll \mu^2$, thus recovering the gapless scalar mode from Eq.~\eqref{massless-decoupled-mode}. Keeping instead $b\,,\,d$ finite, the above equation has the form:
\begin{eqnarray} \label{omega2-of-p}
    \omega_{-}^2({\bf p}) &=& m^2 + \alpha^2({\bf p}) \, {\bf p}^2 \, ,
\end{eqnarray}
where we have defined the momentum-dependent parameter $\alpha^2({\bf p})$ as
\begin{eqnarray} \label{alpha2}
    \alpha^2({\bf p}) = \frac{1}{2} + \tilde{c}_2 \left( - 1 + \frac{{\bf p}^2}{2m^2} \right) + \tilde{c}_3 \left( 1 + \frac{{\bf p}^2}{2m^2} \right) \, ,
\end{eqnarray}
with $\tilde{c}_{2,3} \equiv c_{2,3}/(b+d)$. 

One would be tempted to use the coefficient of the ${\bf p}^2$ term in Eq.~\eqref{omega2-of-p}, namely $\alpha^2({\bf p})$, as a notion of velocity. However, as we will discuss in the next section, this requires some care due to the presence of a gap. After clarifying this, we will use the dispersion relation \eqref{omega2-of-p} to recover the known bounds in Sec.~\ref{sec:bounds}.

\section{IR kinematics and the acoustic metric} \label{sec:acoustic}

In order to study the propagation of the low-energy modes in this theory, it is useful to turn to the effective description given by the acoustic metric. 
The purpose of this Section is to identify IR kinematic quantities that are suitable to describe gapped modes, and that allow to synthetically express the known analyticity bounds. As we will discuss in Sec.~\ref{sec:bounds}, the usual notions of phase velocity and group/ray velocity, see e.g. \cite{Boillat:1969slt,Boillat:1970gw,Landau}, do not lead to a momentum-independent expression of the bounds.
Instead, as we will see, a natural notion of velocity related to both phase and group velocity, and derived from the momentum of perturbations, encodes the known analyticity bounds through a simple, momentum-independent condition.
We now proceed to summarize relevant information to be used in Sec.~\ref{sec:bounds} for discussing the IR characterization of the bounds.

Quite generally, the equations of motion for linearized wave-like excitations $\delta\chi\sim e^{ip_\mu x^\mu}$ of a field $\chi$ 
can be studied in terms of the excitation's momentum $p_\mu$,  
obtaining an on-shell condition of the form:
\begin{eqnarray}\label{eq:Zdef}
    Z^{\mu\nu}(p) p_\mu p_\nu + C^2 = 0 \,,
\end{eqnarray}
where $Z^{\mu\nu}$ is known as the acoustic metric, which we will introduce in detail below, see Sec.~\ref{sec:nondisp}, and $C^2$ is a constant that is associated with the gap of the $\delta\chi$ excitations. 

In the case of an isotropic medium, this equation can be solved by a single dispersion relation for the components of $p_\mu=(-\omega,{\bf p})$:  $\omega=\omega({\bf p})$. From the dispersion relation $\omega({\bf{p}})$, two {\it a priori} different notions of velocity are usually defined\footnote{The appropriate definitions in the case of anisotropic media are discussed e.g. \cite{Sawicki:2024ryt}.}:
\begin{eqnarray}\label{vp}
    |{\bf v}_p| &=& \frac{\omega({\bf p})}{|{\bf p}|} \, , \quad \text{the phase velocity}\;, \\  \label{vg}
    {\bf v}_g &=& \frac{\partial \omega({\bf p})}{\partial {\bf p}} \, , \quad \text{the group velocity}\;.
\end{eqnarray}
The first one describes the propagation of constant phase surfaces of the wave, while the second is commonly associated with the motion of a wave packet. In the geometrical optics regime, the latter also corresponds to the velocity of particle-like trajectories/rays. Formally, the support in spacetime of the retarded Green's function is ultimately given by the front/signal velocity, i.e. the limiting phase velocity at infinite momentum \cite{Diener:1996mj} \footnote{Though typically inaccessible from the low-energy EFT.}. Both notions of velocity defined above can also be obtained from properly defined Lorentz-covariant velocity vectors $U_p^\mu$ and $U_g^\mu$ for the phase and group velocity respectively:
\begin{align}\label{degvpvg}
    {\bf v}_p=\frac{{\bf U}_p}{U_p^0}\;,\quad{\bf v}_g=\frac{{\bf U}_g}{U_g^0}\;.
\end{align}
The Lorentz-covariant phase velocity $U_p^\mu$ can be defined by
\begin{eqnarray}
    U_p^{\mu} p_\mu = 0 
\end{eqnarray}
up to a normalization, while the Lorentz-covariant group velocity $U_g^\mu$ is given by
\begin{eqnarray}
    U_g^{\mu} = \frac{1}{2} \frac{\partial}{\partial p_\mu} \left( Z^{\alpha\beta}(p) p_\alpha p_\beta + C^2 \right) \, .
\end{eqnarray}
This is consistent with Eq.~\eqref{degvpvg} because the equation of motion (EoM) implies 
\begin{align}\frac{\partial}{\partial {\bf p}_i}(Z^{\mu\nu}(p)p_\mu p_\nu)+\frac{\partial}{\partial\omega}(Z^{\mu\nu}(p)p_\mu p_\nu)\frac{\partial\omega({\bf p})}{\partial{\bf p}_i}=-\frac{d}{d{\bf p}_i}C^2=0\;,\end{align} 
and we have defined $\omega=-p_0$.

By Eq.~\eqref{eq:Zdef}, we can recognize that $Z^{\mu\nu}$ defines the soundcone that characterizes on-shell excitations. 
For this reason, it is appropriate to treat $Z^{\mu\nu}$ as a metric for the momentum co-vector $p_\mu$, see also Eq.~\eqref{linearized-EoM} and the related discussion. 
Then it is natural to define an acoustic vector in terms of $Z^{\mu\nu}$ and $p_\mu$:
\begin{eqnarray}\label{eq:N}
    N^{\mu} = Z^{\mu\nu}(p) p_\nu \, ,
\end{eqnarray}
which allows one to write the on-shell condition as
\begin{eqnarray}
    N^{\mu} p_\mu = - C^2 \, .
\end{eqnarray}
The acoustic vector $N^\mu$ often coincides, up to a normalization, with either or both the phase velocity and the group velocity vectors. For instance, up to a normalization factor, we have
\begin{align}\label{eq:conditionsV}\begin{split}
    N^\mu =& \,U_p^{\mu} \; , \qquad \text{for $C = 0$} \;,\\
    N^\mu =& \,U_g^{\mu} \; , \qquad \text{for $\frac{\partial Z^{\alpha\beta}}{\partial p_\mu} = 0$} \;.\end{split}
\end{align}
The three velocities coincide when there is no dispersivity in the system, i.e. when ${\bf v}_p$ and ${\bf v}_g$ are momentum independent: $N^\mu = U_p^{\mu} = U_g^{\mu}$.
Conversely, in the general situation when neither of the conditions of Eq.~\eqref{eq:conditionsV} hold, all three velocity vectors are different $N^\mu \neq U_p^{\mu} \neq U_g^{\mu}$. Nevertheless, the acoustic vector $N^\mu$ still contains a valid notion of velocity:
\begin{eqnarray}\label{contra-vel}
    {\bf v} &=& \frac{{\bf N}}{N^0} \, .
\end{eqnarray}

In what follows we discuss the acoustic metric $Z^{\mu\nu}$ in more detail and motivate the use of the vector $N^\mu$ as a notion of velocity with which to express the known analyticity bounds. The reader interested only on the recovery of the bounds may skip the rest of this Section and instead go directly to Sec.~\ref{sec:bounds}. For a more detailed account of this formalism see Ref.~\cite{Sawicki:2024ryt}.

\subsection{Non-dispersive, gapless case}\label{sec:nondisp}

We now discuss the cases in which the system has second order equations for perturbations with no gap, and introduce the acoustic metric $Z^{\mu\nu}$ in detail. These assumptions amount to having $C=0$ and $\partial Z^{\alpha\beta}/\partial p_\mu=0$, meaning that the system has no dispersivity. For simplicity, we also consider an isotropic medium.
We begin by considering a theory with gapless excitations $\delta\chi$ described by a quadratic Lagrangian of the form
\begin{eqnarray}
    \mathcal{L}_2 = - \frac{1}{2} \sqrt{-S} Z^{\mu\nu} \partial_\mu \delta \chi \partial_\nu \delta \chi \, ,
\end{eqnarray}
where in this paragraph we assume $Z^{\mu\nu}$ to contain no differential operators, meaning that the theory has second order equations of motion. Here, for convenience, we have introduced $S_{\mu\nu}$ to be the inverse of $Z^{\mu\nu}$, and defined $S = \det(S_{\mu\nu})$. In this case the associated linearized EoM can be written as
\begin{eqnarray} \label{linearized-EoM}
    \frac{1}{\sqrt{-S}} \partial_\mu \left( \sqrt{-S} Z^{\mu\nu} \partial_\nu \delta \chi \right) = \bar{\square} \delta \chi = 0 \, ,
\end{eqnarray}
with $\bar{\square} = Z^{\mu\nu} \bar{\nabla}_\mu \bar{\nabla}_\nu$ and $\bar{\nabla}_\mu$ is the covariant derivative metric-compatible with $Z$ and $S$. This is simply the wave-equation on the nontrivial background given by the acoustic metric.

In a general case in which $Z^{\mu\nu}(x)$ has a non-trivial space-time dependence, e.g. the background is not homogeneous, we study this equation in the Eikonal approximation, with an ansatz $\delta \chi = A(x) \, e^{i \mathcal{S}(x)/\epsilon}$, and expanding for $\epsilon \to 0$. This amounts to consider a rapidly varying phase $\mathcal{S}(x)$ and a slowly varying amplitude $A(x)$, appropriate for the geometric optics regime in which the wavelength of the excitation $\delta\chi$ is much shorter than the typical scales of the background inhomogeneities. In this limit, one can show that at leading order in $1/\epsilon$ Eq.~\eqref{linearized-EoM} corresponds to the on-shell condition:
\begin{eqnarray} 
    Z^{\mu\nu}(x) p_\mu p_\nu = 0 \, ,
\end{eqnarray}
where now $p_\mu = \partial_\mu \mathcal{S}$ is the momentum co-vector of the excitation. Since by assumption the theory is second order, $Z^{\mu\nu}$ does not itself depend on $p_\mu$.

We may construct a geometrical picture of how the excitations propagate in this regime by deriving a corresponding geodesic equation associated to $Z^{\mu\nu}$. Acting with Z-compatible vector derivative $\bar{\nabla}_\alpha$ on the above on-shell condition, we get
\begin{eqnarray}\label{eq:paral}
    Z^{\mu\nu} p_\nu \bar{\nabla}_\mu p_\alpha = 0 \, ,
\end{eqnarray}
where we used the fact that $p_\mu$ is the gradient of a scalar quantity, the phase $\mathcal{S}$, and therefore $\bar{\nabla}_\alpha p_\mu = \bar{\nabla}_\mu p_\alpha$. Finally, as in Eq.~\eqref{eq:N}, we can define the acoustic vector $N^\mu$ by raising the index of the momentum co-vector $p_\mu$ with $Z^{\nu\alpha}$,
\begin{eqnarray}
    N^\mu = Z^{\mu\nu} p_\nu \, .
\end{eqnarray}
This leads to the following geodesic equation:
\begin{eqnarray} \label{geodesic-N}
    N^\mu \bar{\nabla}_\mu N^\nu = 0 \, .
\end{eqnarray}
In light of this, Eq.~\eqref{eq:paral} can be rephrased as the parallel transport of energy-momentum along the $N^\mu$ geodesic, $N^\mu \bar{\nabla}_\mu p_\nu = 0 $. Physically, the acoustic vector $N^\mu$ tracks the flow of energy-momentum $p_\mu$. 
Note that, by definition of $N^\mu$, the EoM can be rewritten as
\begin{eqnarray}
    S_{\mu\nu} N^\mu N^\nu = 0 \, .
\end{eqnarray}
In conclusion, the equations of motion in the gapless and dispersionless case imply that the acoustic vector $N^\mu$ follows null geodesics of $S_{\mu\nu}$. This defines the trajectories, also known as ``rays", describing propagation of the wavefronts in the geometric optics approximation:
\begin{eqnarray} \label{eq:phase-surf}
    N^\mu p_\mu = N^\mu \partial_\mu \mathcal{S} = 0 \quad , \quad \text{for gapless, dispersionless modes.}
\end{eqnarray}
This equation means that the vector $N^\mu$ defines surfaces of constant phase.
In fact, as we discussed, in the present nondisperive cases $N^\mu$ coincides with the phase velocity as well as the group velocity vectors. Therefore, the corresponding velocity, Eq.~\eqref{contra-vel},
\begin{eqnarray}\nonumber
    {\bf v} = \frac{{\bf N}}{N^0} \, ,
\end{eqnarray}
coincides with both phase and group velocity, defined in Eqs.~\eqref{vp} and~\eqref{vg}. 

The description of kinematics in terms of the acoustic metric allows one to retrieve known results (see e.g. Appendix~\ref{app:PX}), while enabling a geometrical interpretation of more complicated scenarios. As we will see, the velocity defined through the acoustic vector $N^\mu$ will remain a relevant quantity in dispersive cases, when the propagation is affected by higher derivatives as well as by a gap, and the equivalence between the different notions of velocity is no longer valid.

\subsection{Gapped and dispersive systems} \label{sec:beyond-no-gap}

For our particular purpose of studying the gapped mode \eqref{omega2-of-p} in the low-momentum regime, we need to depart from the Eikonal approximation of the previous section and consider the regime in which momenta are not necessarily large enough to allow us to neglect the terms in the EoM with less derivatives. To keep the discussion simple, we consider a space-time independent acoustic metric and gap term. This is exactly the case for the model discussed in Sec.~\ref{sec:superf}, since there we have $\partial_\rho Z^{\mu\nu} = 0$; see Eq.~\eqref{lagrangian2}. This means that we can use a plane-wave ansatz $\delta \chi = A \, e^{ip_\mu x^\mu}$, with $A$ and $p_\mu$ constants. We thus obtain
\begin{eqnarray} 
    Z^{\mu\nu} p_\mu p_\nu + C^2 = 0 \, ,
\end{eqnarray}
which implies a gap $m^2 = C^2/|Z^{00}|$. 

Following the same steps as for the nondispersive case, one can still define the acoustic vector $N^\mu = Z^{\mu\nu} p_\nu$ and see that the EoM sets this vector as $S$-timelike:
\begin{eqnarray}
    S_{\mu\nu} N^\mu N^\nu = - C^2 \, .
\end{eqnarray}
Of course, the vector $N^\mu$ still follows the geodesic equation, Eq.~\eqref{geodesic-N}, although this is now trivial, as we are restricting to a flat acoustic metric, $\partial_\rho Z^{\mu\nu} = 0$.
Again, the velocity of rays can be defined as before through the acoustic vector $N^\mu$ by Eq.~\eqref{contra-vel}. This velocity will still coincide with the group velocity, as long as the acoustic metric is momentum independent, i.e., $\partial Z^{\alpha\beta}/\partial p_\mu=0$. If instead the metric depends on the momentum, we will have that the two notions of velocity are different:
\begin{align}
    U_g^{\mu} = N^\mu + \frac{1}{2} \frac{\partial Z^{\alpha\beta}(p)}{\partial p_\mu} p_\alpha p_\beta \, .
\end{align}
As a consequence of the presence of a gap, we now have $N^\mu p_\mu =-C^2\neq0 $, implying that in this case the speed defined through $N^\mu$ is no longer the phase velocity, as expected for a dispersive case.

As a short and rather trivial example, we can consider a Lorentz invariant acoustic metric, i.e. $Z^{\mu\nu} = \eta^{\mu\nu}$ and $S_{\mu\nu} = \eta_{\mu\nu}$, in the presence of a gap $C^2$ (invariant mass $m^2$ in this case). The previous two equations give the dispersion relation:
\begin{eqnarray}
    \omega^2({\bf p}) = m^2 + {\bf p}^2 \, .
\end{eqnarray}
Notice that with this, the phase velocity ${\bf v}_p^2 = \omega^2({\bf p})/{\bf p}^2$ is divergent for ${\bf p} \to 0$. 
Instead, the equation
\begin{eqnarray}
    (N^0)^2 = m^2 + {\bf N}^2 
\end{eqnarray}
allows one to compute the velocity ${\bf v}$ from the acoustic vector $N^\mu$, Eq.~\eqref{contra-vel}, which is well-defined at zero momentum.
Given the form of the acoustic metric in this example, it takes the form
\begin{align}
    {\bf v}^2=\frac{{\bf N}^2}{(N^0)^2}=\frac{{\bf p}^2}{m^2+{\bf p}^2}\leq 1\;,
\end{align}
which coincides in this case with the group velocity ${\bf v}_g^2$, as discussed above. 

While in this section we have focused in the case relevant to our purposes, of a space-time independent acoustic metric, let us remark, that in the case of a space-time dependent acoustic metric that depends on $p$, a generalization of Riemannian geometry is required and the relation between geodesics and auto-parallel curves becomes more involved \cite{miron2002geometry}.

\section{Bounds in terms of the IR kinematics}\label{sec:bounds}

\subsection{Conformal superfluid in D=3}

We now return our focus to the low-energy ($\omega \ll \mu$) mode $\omega_{-}^2({\bf p})$, whose acoustic metric will be momentum-dependent on account of the $c_{2,3}$ contributions.

From the dispersion relation \eqref{omega2-of-p}, we can read off the acoustic metric\footnote{Actually, we can only read it up to a conformal factor, which for our discussion is irrelevant so we choose $Z^{00} = 1$.}
\begin{eqnarray}
    Z^{\mu\nu}(p) \, p_\mu p_\nu + m^2 = 0 \, ,
\end{eqnarray}
with $Z^{\mu\nu} = diag( -1, \alpha^2, \alpha^2)$. An important remark here is that we are treating the contributions from HD operators, responsible for the ${\bf p}$-dependence in $\alpha^2({\bf p})$, as perturbative within the regime of validity of the EFT, implicitly giving a condition on $c_{2,3}$ and $p_\mu$. This prompts us to define a \mbox{$p_\mu$-dependent} acoustic metric $Z^{\mu\nu}(p)$, which we then otherwise treat as usual. In particular, its inverse is simply $S_{\mu\nu} = diag( -1, 1/\alpha^2, 1/\alpha^2)$ and trading $p_\mu$ for $N^\mu = Z^{\mu\nu}(p) p_\nu$ can always be unambiguously done perturbatively. Rewriting the on-shell condition in terms of $N^\mu$, we get 
\begin{eqnarray}
    - (N^0)^2 + \frac{{\bf N}^2}{\alpha^2} + m^2 = 0 \, ,
\end{eqnarray}
which clearly would give the usual ${\bf v}^2 = {\bf v}_p^2 = \alpha^2$ if there was no gap. Instead, in this case, we have
\begin{eqnarray}
    \left(- \frac{1}{{\bf v}^2} + \frac{1}{\alpha^2} \right) {\bf N}^2 + m^2 = 0 \, ,
\end{eqnarray}
where we used Eq.~\eqref{contra-vel}. It is useful to define the parameter
\begin{eqnarray}\label{vareps}
    \varepsilon \equiv \frac{{\bf p} \cdot {\bf N}}{m^2} 
    = \alpha^2 \frac{ {\bf p}^2}{m^2} \,. 
\end{eqnarray}
Then we find
\begin{eqnarray}
    \left(- \frac{1}{{\bf v}^2} + \frac{1}{\alpha^2} \right) \alpha^2 \varepsilon + 1 = 0 \, ,
\end{eqnarray}
which leads to the following expression for the velocity ${\bf v}$ associated to the acoustic vector $N^\mu$,
\begin{eqnarray} \label{contra-vel-massive}
    {\bf v}^2 = \alpha^2 \frac{\varepsilon}{1 + \varepsilon} \, .
\end{eqnarray}
As anticipated, the gap induced by the coupling between the scalar and gauge modes slows down ${\bf v}$ at low momenta, where it significantly differs from the phase-velocity ${\bf v}_p$. Nevertheless, the two velocities approximately coincide for momenta well above the gap\footnote{Note that $\varepsilon\gg 1$ can be in the regime of validity of the EFT as long as $b+d\gg c_1$, as we have that the mass gap is much below the cutoff of the EFT: $m^2\ll \mu^2$.} ($\varepsilon \gg 1$), i.e. ${\bf v}^2 \simeq \omega_{-}^2({\bf p})/{\bf p}^2 \simeq \alpha^2({\bf p}) $. Moreover, note that the $p$-dependence of the acoustic metric implies that ${\bf v}$ also differs from the group velocity ${\bf v}_g$.

As discussed, we expect the velocity of a gapless mode in a conformal superfluid in $D=3$ to run from the limiting value $c_s^2 = 1/2$ at zero momentum to $1$ at infinite momentum. As usual, we will assume that the running of this speed is strictly positive, corresponding to a stable system. Therefore, to provide a bound on $\alpha^2({\bf p})$, the physically meaningful condition should be to demand that velocities remain below $c_s^2 = 1/2$ for sufficiently low momenta, and below $c=1$ for high momenta. With this in mind, a relevant question is which notion of velocity makes it possible to capture the analyticity bound \eqref{bound-xi} through a sharp, momentum-independent condition. In the following, we show that such a momentum-independent condition can be found for the velocity ${\bf v}$ defined from the acoustic vector $N^\mu$.

According to the previous considerations, we examine a momentum-independent condition on ${\bf v}$, namely,
\begin{eqnarray} \label{bound}
    {\bf v}^2 \leq c_s^2 = \frac{1}{2} \, .
\end{eqnarray}
This can be equivalently written in terms of $\alpha^2({\bf p})$,
\begin{eqnarray} \label{bound-alpha}
    \alpha^2({\bf p}) \leq \frac{1 + \varepsilon}{2\varepsilon} \, .
\end{eqnarray}
This condition should not be imposed for too high momenta, as it is expected that eventually ${\bf v}^2$ will increase toward $1$. Since the coefficients of the EFT are not bounded when the theory is gapless, we know that the maximum momentum should scale like $m$, rather than $\mu$.
For instance, we can stop at $\varepsilon=1$,\footnote{From Eq.~\eqref{vareps}, we can see that momentum ${\bf p}_*$ at which $\varepsilon = 1$ satisfies $\alpha^2({\bf p}_*) {\bf p}_*^2 = m^2$. Since $\alpha^2({\bf p})$ should always be increasing, then we also have $\alpha^2({\bf p}) \geq 1/2$ for all ${\bf p}$. Therefore, the scale ${\bf p}_*$ lies in somewhere in the range $m^2 \leq {\bf p}_*^2 \leq 2 m^2$. When $\alpha^2({\bf p})$ saturates the bound \eqref{bound}, i.e. $\alpha^2_\text{max} = 1$ at $\epsilon = 1$, we have ${\bf p}_*^2 = m^2$.}  where the momentum starts being important with respect to the mass in the energy balance of Eq.~\eqref{omega2-of-p}.
As a matter of fact, we find that for $\varepsilon>1$, 
it will be $\alpha^2 < 1$, which limits the speed of excitations to be strictly below that of a Lorentz invariant particle at the same mass.
\footnote{For a free, massive Lorentz-invariant particle, ${\bf v}^2 = {\bf v}_g^2 \geq 1/2$ for $p^2 \geq m^2$.}  
 
As we will see, these natural considerations lead to reproducing the analyticity bounds of Eq.~\eqref{bound-xi}, derived in \cite{Creminelli:2022onn}. Recalling the definition for $\alpha^2$ from Eq.~\eqref{omega2-of-p} and using Eq.~\eqref{bound-alpha}, we may now explicitly write Eq.~\eqref{bound} as
\begin{eqnarray} \label{bound-expl}
    \tilde{c}_2 (\varepsilon - 1) + \tilde{c}_3 (\varepsilon + 1) \leq \frac{1}{2\varepsilon} \, ,
\end{eqnarray}
with $\varepsilon \in [0,1]$. The optimal choice of $\varepsilon$ actually depends on the values of $\tilde{c}_2$ and $\tilde{c}_3$. In this form our result can be readily compared to the known analyticity bound \eqref{bound-xi}. For this purpose, it is useful to change variables by defining $\frac{(1 - \varepsilon)}{(1 + \varepsilon)} \equiv 1 - \xi^2$. For the range $\varepsilon \in [0,1]$ that we consider, the new variable is also $\xi \in [0,1]$, while \eqref{bound-expl} can be rewritten as
\begin{eqnarray}
    \tilde{c}_2 \left( 1 - \xi^2 \right) - \tilde{c}_3 \geq - \frac{\left( 1 - \frac{\xi^2}{2} \right)^2}{\xi^2} \, .
\end{eqnarray}
This confirms that the bounds derived in Ref.~\cite{Creminelli:2022onn} [see Eq.~(58) therein], are in fact a sharp statement about ${\bf v}$, that is, about the acoustic vector $N^\mu$.

Notice that the same bound cannot be expressed in terms of a momentum-independent condition on either the group or the phase velocity. In fact, as discussed in the previous Section, we know that in this situation ${\bf v}$ is different from both the phase velocity ${\bf v}_p$ and the group velocity ${\bf v}_g$. Therefore, the bound cannot constrain either of these velocities with a momentum-independent condition. This can be confirmed by direct computation,
\begin{eqnarray}
    {\bf v}_p^2 &=& \frac{m^2}{{\bf p}^2} + \left[ \frac{1}{2} + \bar{c}_2 \left( -1 + \frac{{\bf p}^2}{2m^2} \right) + \bar{c}_3 \left( 1 + \frac{{\bf p}^2}{2m^2} \right) \right] \, , \\
    {\bf v}_g &=& \left[ \frac{1}{2} + \bar{c}_2 \left( -1 + \frac{{\bf p}^2}{m^2} \right) + \bar{c}_3 \left( 1 + \frac{{\bf p}^2}{m^2} \right) \right] \frac{{\bf p}}{\omega} \, .
\end{eqnarray}

It is also interesting to consider the limiting case where $b+d \to \infty$, where the gauge field decouples. Recalling that $\tilde{c}_{2,3} = c_{2,3}/(b+d)$ and taking this limit in Eq.~\eqref{bound-expl} at fixed $\varepsilon$, gives as a result a trivialization of the bound. In other words, we recover the expected outcome that for the pure scalar theory we are unable to extract a meaningful bound, as discussed in Sec.~\ref{sec:scalar-mode}. From the point of view of our argument, this can be understood as a consequence of the vanishing gap for the scalar field, which implies that Eq.~\eqref{bound} can only be applied at zero momentum, giving a trivial bound.

\subsection{Interpretation}

We have seen that the analyticity bounds can be transparently phrased as a momentum-independent condition \eqref{bound} on the velocity ${\bf v}$ corresponding to the acoustic vector $N^\mu$, but due to dispersivity and the presence of a gap, this does not translate into an equally sharp bound on neither the phase velocity ${\bf v}_p$ nor the group velocity ${\bf v}_g$. It is useful then to see this condition directly as a constraint on the momentum of the gapped mode. Indeed, one may rewrite the bound of Eq.~\eqref{bound} as
\begin{eqnarray} \label{cone-cond}
    N^\mu \bar{p}_\mu \leq 0 \,,
\end{eqnarray}
where $\bar{p}_\mu = \left( c_s |{\bf p}| , {\bf p} \right)$ is a fiducial on-shell momentum of a gapless particle with speed $c_s^2=1/2$, and ${\bf p}$ is the momentum of the gapped excitation (at which $N^\mu$ is evaluated). 
This means that the acoustic vector $N^\mu$ lies inside of the sound-cone of the gapless excitation. This soundcone can be defined through $\bar{N}^\mu = \bar{Z}^{\mu\nu} \bar{p}_\nu$, with $\bar{Z}^{\mu\nu} = diag(-1 , c_s^2, c_s^2)$, as this vector satisfies: $\bar{N}^\mu \bar{p}_\mu = 0$.

As discussed, the same condition cannot be phrased in a momentum independent way in terms of either the phase velocity $U_p^\mu$, the group velocity $U_g^\mu$, or correspondingly the momentum co-vector $p_\mu$, as the bound Eq.~\eqref{bound} would read $\bar{p}_\mu Z^{\mu\nu}(p) p_\nu\leq 0$, which is a condition on momentum dependent combinations of the components of $p_\mu$.

\subsection{Connection with $\langle J J\rangle$ correlators}\label{sec:jj}

In Ref.~\cite{Creminelli:2022onn} analyticity bounds were derived for the conformal superfluid model studied above, by studying a Green's function associated with the correlator of conserved currents $J^\mu$, computed in the case where the gauge field remains non-dynamical. The bounds obtained in this case coincide with our result in Eq.~\eqref{bound}, which instead we obtained by studying the low-energy theory as-is, allowing the gauge field to be dynamical and therefore kinetically mix with the scalar. 

To see the connection between these two approaches, consider the path integral
\begin{eqnarray}
    Z = \int \mathcal{D} \pi \, e^{i \int d^3 \! x \, \mathcal{L}(\pi, A_\mu)} \,.
\end{eqnarray}
Focusing on the quadratic Lagrangian of Eq.~\eqref{lagrangian}, which we now parameterize as
\begin{eqnarray}
    \mathcal{L}_2(\pi, A_\mu) = \frac{1}{2} \left[ \int d^3 \! x' \, \pi(x) G^{-1}_{\pi\pi}(x,x') \pi(x') + 2 A_\mu(x) J^\mu(x) + \int d^3 \! x' A_\mu(x) [G^{-1}]^{\mu\nu}(x,x') A_\nu(x') \right] \, , \notag \\
\end{eqnarray}
we can compute correlators by taking functional derivatives. In particular, taking two functional derivatives with respect to $A_\mu$ we get
\begin{eqnarray} \label{2nd-var-Z}
    \frac{1}{Z} \frac{\delta^2 Z}{\delta A_\mu(x) \delta A_\nu(x')} \Bigg|_{A=0} &=& \frac{1}{Z} \int \mathcal{D} \pi \, \left[ - J^\mu(x) J^\nu(x') + i [G^{-1}]^{\mu\nu}(x,x') \right] \, e^{i \int d^3 \! x \, \mathcal{L}_2(\pi, A_\mu)} \notag \\
    &=& - \langle 0 | T \left\{ J^\mu(x) J^\nu(x') \right\} | 0 \rangle + i [G^{-1}]^{\mu\nu}(x,x') \, ,
\end{eqnarray}
while, if we go ahead and integrate out the scalar $\pi$, we can compute the Gaussian functional integral explicitly and obtain
\begin{eqnarray} \label{integrated-out-Z}
    Z = N \, \exp\left({\frac{i}{2} \int_{x,x'} \!\!\!\!\!\!\!\! A_\mu(x) \left\{ [G^{-1}]^{\mu\nu}(x,x') - \int _{x_1,x_2} \!\!\!\!\!\!\!\!\!\!\, \tilde{J}^\mu(x,x_1) G_{\pi\pi}(x_1,x_2) \tilde{J}^\nu(x_2,x') \right\} A_\nu(x') \, } \right) \, , \notag \\ 
\end{eqnarray}
where we rewrite the linear-in-$A_\mu$ term in the quadratic Lagrangian as
\begin{eqnarray}
    A_\mu(x) J^\mu(x) = \int d^3 \! x' \, A_\mu(x) \tilde{J}^\mu(x,x') \pi(x') \, .
\end{eqnarray}
With this result we can immediately find that the correlator of conserved currents is
\begin{eqnarray}
    \langle 0 | T \left\{ J^\mu(x) J^\nu(x') \right\} | 0 \rangle = i \int d^3 \! x_1 d^3 \! x_2 \, \tilde{J}^\mu(x,x_1) G_{\pi\pi}(x_1,x_2) \tilde{J}^\nu(x_2,x') \, .
\end{eqnarray}
\sloppy Notice that while this correlator is only conserved up to a contact term 
$p_\mu \langle 0 | T \left\{ J^\mu(x) J^\nu(x') \right\} | 0 \rangle \sim D^\nu(x) \delta(x-x')$, where $D^\nu(x)$ is local a differential operator, its retarded version is conserved exactly.

At this point it is also clear that, if the gauge field $A_\mu$ were to be promoted to a dynamical field, its effective kinetic term after integrating out the scalar would be the quantity in curly brackets in the right-hand-side of Eq.~\eqref{integrated-out-Z}, which is also the last expression in Eq.~\eqref{2nd-var-Z}. This shows that the $\langle JJ \rangle$ correlator contains the information of the kinetic mixing between the scalar and the gauge field. Indeed, this correlator can potentially shift the pole of the $A_\mu$ propagator, which determines the kinematics of the gauge field excitations. In Fourier space
\begin{eqnarray}\label{eq:det}
    \det\left( [G^{-1}]^{\mu\nu}(\omega, {\bf p}) + \hat{J}^\mu(-\omega, -{\bf p}) G_{\pi\pi}(\omega, {\bf p}) \hat{J}^\nu(\omega, {\bf p}) \right) = 0 \, ,
\end{eqnarray}
where $\hat{J}^\mu(\omega, {\bf p})$ is $-i$ times the Fourier transform of $\tilde{J}^\nu(x,x')$, with components given explicitly in Eqs.~\eqref{hatJ0} and \eqref{hatJi}. Upon explicit computation, using a gauge fixing, the above can be seen to give the same condition as Eq.~\eqref{pi-A0-mode-eq}. This happens because the eigenmode of the kinetic matrix \eqref{kinetic-matrix} mixes with $A_\mu$ and changes its propagation. Therefore the solution to Eq.~\eqref{eq:det} is precisely the mode $\omega_{-}^2({\bf p})$ given in Eq.~\eqref{omega2-of-p}, which was central to the derivation of the bounds derived in Ref.~\cite{Creminelli:2022onn} from the $\langle JJ \rangle$ correlator.

\subsection{Example of UV completion}\label{sec:UV-compl}

In this Section we outline a specific example of a UV completion of the conformal superfluid in $D=3$ in the absence of the gauge field, consisting of a conformal complex scalar field, as also studied in Ref.~\cite{Creminelli:2022onn}. This provides an explicit example of the momentum dependence of the velocity $\bf{v}$ obtained from the acoustic vector $N^\mu$ of the gapless mode, which transitions from $1/2$ at zero momentum to $1$ at momenta much larger than the scale of symmetry breaking $\mu$.

The Lagrangian for this UV-complete model is given by \cite{Badel:2019khk}
\begin{eqnarray}
    S_\mathrm{UV} &=& \sqrt{-g} \left( -|\partial \phi|^2 - \lambda |\phi|^6 - \frac{1}{8} R |\phi|^2 \right) \notag \\
    &=& \sqrt{-g} \left( -(\partial \rho)^2 - \rho^2 (\partial \theta)^2 - \lambda \rho^6 - \frac{1}{8} R \rho^2 \right) \, ,
\end{eqnarray}
where $\phi = \rho \, e^{i\theta}$ and $\lambda > 0$. Upon integrating out the modulus field $\rho$ at tree level, the EFT action from Eq.~\eqref{lagrangian} is obtained with coefficients \cite{Creminelli:2022onn}
\begin{eqnarray}
    c_1 = \frac{4}{\sqrt{3\lambda}} \, , \qquad c_2 = \frac{1}{8\sqrt{3\lambda}} \, , \quad c_3 = b = d = 0 \, ,
\end{eqnarray}
plus higher-dimensional operators. Coupling to an external gauge field $A_\mu$ can be done by introducing a gauge-covariant derivative $\partial_\mu \to \mathcal{D}_\mu$. Expanding around a state with chemical potential $\mu$, $\theta = \mu t + \frac{\pi_c}{\sqrt{2} \bar{\rho}}$ and $\rho = \bar{\rho} + \frac{\delta \rho_c}{\sqrt{2}}$, where $\bar{\rho} \simeq \sqrt{\mu}/(3\lambda)^{1/4}$, the quadratic Lagrangian for the fluctuations $\pi_c$ and $\delta \rho_c$ has the following kinetic matrix in Fourier space
\begin{eqnarray}
    \mathcal{K}_{IJ}(\omega, {\bf p}) = 
    \begin{pmatrix}
        \omega^2 - {\bf p}^2 & 2i \mu \omega \\
        -2i \mu \omega & \quad\omega^2 - {\bf p}^2 - 4\mu^2 \\
    \end{pmatrix} \, ,
\end{eqnarray}
where $I,J = \{\rho, \theta\}$. The corresponding eigenmodes $\omega({\bf p})$ are obtained from $\det(\mathcal{K}) = 0$,
\begin{eqnarray} \label{gapless-mode-UV-compl}
    \omega_{\phi\pm}^2({\bf p}) = {\bf p}^2 + 4\mu^2 \left( 1 \pm \sqrt{1 + \frac{{\bf p}^2}{4\mu^2}} \right) \, .
\end{eqnarray}
The minus branch gives rise to a gapless mode, the superfluid phonon, while the plus branch describes a massive mode with $M^2 = 8\mu^2$. Within the regime of validity of the EFT, only the former can be excited. It is instructive to examine its dispersion relation in both the regimes of low and high momentum $|{\bf p}|$ with respect to $\mu$,
\begin{eqnarray}
    \omega_{\phi-}^2({\bf p}) \simeq 
    \begin{cases}
        \frac{{\bf p}^2}{2} & \qquad |{\bf p}| \ll \mu \, , \\
        {\bf p}^2 & \qquad |{\bf p}| \gg \mu \, .
    \end{cases}
\end{eqnarray}
The low-energy side of this should be compared to Eq.~\eqref{massless-decoupled-mode}, i.e. the gapless mode of the EFT without any dynamical gauge field. 

Since this mode is gapless, the phase velocity and the velocity from the acoustic vector $N^\mu$ coincide ${\bf v}^2 = {\bf v}_p^2=\omega_{\phi-}^2({\bf p})/{\bf p}^2$. The above example exhibits how the velocity ${\bf v}^2$ of this mode interpolates between the soundspeed of a conformal superfluid in $D=3$, $c_s^2 = 1/2$, at low energies when the Lorentz invariance and the conformal invariance are spontaneously broken by the finite chemical potential $\mu$, and the speed of light $c^2=1$ in the unbroken phase at high energies.

\section{Discussion}

In this work we have approached the question of what is the IR characterization of analyticity/positivity bounds in theories with spontaneously broken Lorentz symmetry. We have expressed the known bounds in terms of a momentum-independent condition on the kinematics of low energy modes in the EFT, specifically through the acoustic vector $N^\mu$ and the associated velocity ${\bf v}$. We studied this in the context of an EFT for a conformal superfluid in $D=3$, presented in Sec.~\ref{sec:superf}, studying an IR model in which the $U(1)$ symmetry group is gauged and the superfluid is coupled to a gauge field. In this model the EFT has both a mass gap and HD operators, both contributing to the presence of dispersivity. In Sec.~\ref{sec:bounds} we recovered the known analyticity bounds for this theory by requiring that the velocity ${\bf v}$ from the acoustic vector $N^\mu$ does not grow too fast for low momenta. Specifically, we have imposed that it does not grow beyond the speed of gapless excitations, $c_s^2=1/2$, for momenta below the mass gap; see Eq.~\eqref{bound}. In doing so, we also learned that the bound is not expressible as momentum-independent condition on neither phase nor group velocities.

Technically, the the fact that the bound of \cite{Creminelli:2022onn} represents a statement about the kinematics of the gapped modes of the theory can be traced back to the close connection between correlators of conserved currents and propagators of the low energy modes, which we discussed in Sec.~\ref{sec:jj}.

In our work, we considered the IR kinematics of the gauge field excitations. Conservatively, this gauge field can be thought of as an auxiliary variable that is helpful in order to study the interactions, similarly to the approach of \cite{Creminelli:2022onn}. It would be interesting to investigate whether this system can also be considered as a physical, stable, low-energy phase of a conformally invariant UV theory. [See e.g., \cite{Kaplan:2023fbl,Seiberg:2024wgj,Berezhiani:2024boz} for recent progress on gauge theories at finite charge density.] In this regard, it would be non-trivial to show that one can have a phase with vanishing electric field, even using a background charge density. Another interesting point is that the EFT we study only has gapped modes. It will be useful to explain this either in terms of analogy with plasma physics, or in terms of the algebra of broken generators, see e.g., \cite{Nicolis:2012vf}.

In conclusion, we have learned that the requirement that modes below the mass gap threshold should be slower, in a specific sense, than the gapless modes, leads to meaningful bounds on the IR coefficients. Equivalently, the energy is bounded by a combination of the mass and the energy of a gapless mode with the same spatial momentum. This is the information contained within the known analyticity bounds. Complementarily, the fact that these bounds do not express sharp statements in terms of neither phase nor group velocities of the any IR modes of the theory indicates that causality has a nuanced character in this system. Our work moves in the direction of developing a simple way to characterize the UV properties of nature from IR quantities, even when the systems observed cannot be described within the most conservative assumptions.

As a direct extension of our work, one might consider bounds derived for conformal superfluids in Ref.~\cite{Creminelli:2022onn}, for other higher-order operators that we have not studied here. The analyticity bounds for these operators were derived studying two-point correlators of the stress-energy tensor. On the IR side, learning from our results one may expect that the coupling to gravity will be important in this case. Considering the propagation of modes which mix the conformal scalar and gravitational excitations may provide the necessary handle to recover the bound on the coefficients of these irrelevant operators from the IR. We may expect this to extend to similar but more general cases in which a symmetry can be gauged, though a full rigorous derivation of this relation is beyond the scope of this paper.

The extension of these results to other systems without Lorentz invariance requires further analysis, as many open questions remain. It would be premature to state here that analyticity bounds will in general be expressible in the IR as a momentum-independent condition on the velocity defined from the acoustic vector $N^\mu$. However, note that for gapless theories this requirement is the same as finding a momentum-independent bound on the phase velocity, which is known to reproduce analyticity bounds in several cases. 
The presence of a gap here establishes a window in momenta where the bound is effectively applied, but this may be a peculiarity that does not generalize beyond the conformal superfluid in $d=3$. Still, our result confirms the expectation that consistent interactions slow down excitations with respect to the free theory, which may still be the case for more general systems. What is clear from our results is that analyticity bounds do not always take the form of a simple comparison between the speed of IR excitations and the speed of light. Instead, the acoustic metric allows to capture more nuanced properties of the IR kinematics that synthetically express analyticity bounds.

In the future, it would also be interesting to turn our attention toward other domains where these ideas might be applicable. Indeed, while the conformal superfluid EFT that we studied here is interesting in its own right, it can also be considered as a proxy for more complicated theories in which Lorentz invariance is spontaneously broken. Such is the case of cosmological theories of modified gravity, our original motivation.

\acknowledgments
The authors would like to thank Gabriel Cuomo, Ignacy Sawicki, Georg Trenkler and Enrico Trincherini for useful discussions, and Alessandro Podo for discussions and comments on the earlier version of this work. The work of F.S. was supported by the Simons Investigator Award No. 827042 (P.I.: Surjeet Rajendran) and the Simons Investigator Award No. 144924 (P.I.: David E. Kaplan). The work of L.G.T. was supported by European Union (Grant No. 101063210).

\appendix

\section{An example -- P(X) theory}\label{app:PX}

To see that working with the geometrical description based in the acoustic metric reproduces the usual known results, let us consider the simple example of a $P(X)$ theory with Lagrangian
\begin{eqnarray}
    \mathcal{L} = -\frac{1}{2} (\partial \phi)^2 + \frac{1}{4} c_2 (\partial \phi)^4 \, .
\end{eqnarray}
As this is a second-order theory without a gap, there is no dispersivity. The corresponding acoustic metric can be obtained by expanding at quadratic order in perturbations around a fixed background, $\phi = \bar{\phi} + \varphi$, and taking the Eikonal limit. It reads as
\begin{eqnarray}\label{Z-PX}
    Z^{\mu\nu} &=& \left ( 1 - c_2 (\partial  {\phi})^2 \right) \eta^{\mu\nu} - 2 c_2 \, \partial^\mu  {\phi} \, \partial^\nu  {\phi} \, ,
\end{eqnarray}
which is generally spatially anisotropic except in the frame where ${\vec{\nabla}} \phi = 0$, which only exists if $(\partial \phi)^2 < 0$. Here for ease of notation we dropped the bar from the background quantities. The corresponding dispersion relation is
\begin{eqnarray}
    \omega^2({\bf p}) = {\bf p}^2 - \frac{2 c_2 (\partial\phi \cdot p)^2}{(1 - c_2 (\partial \phi)^2)} \, ,
\end{eqnarray}
which is gapless. We proceed to compute the inverse of $Z^{\mu\nu}$, easily found to be
\begin{eqnarray}\label{S-PX}
    S_{\mu\nu} &=& \frac{1}{\left ( 1 - c_2 (\partial  {\phi})^2 \right)} \left[ \eta_{\mu\nu} + \frac{2 c_2 \, \partial_\mu  {\phi} \, \partial_\nu  {\phi}}{\left ( 1 - 3 c_2 (\partial  {\phi})^2 \right)} \right] \, .
\end{eqnarray}
This allows one to write the EoM in the geometric optics approximation in terms of the acoustic vector $N^\mu$, obtaining
\begin{eqnarray} \label{N-EoM-PX}
    \eta_{\mu\nu} N^\mu N^\nu + \frac{2 c_2 \, (\partial{\phi} \cdot N)^2}{\left ( 1 - 3 c_2 (\partial  {\phi})^2 \right)} &=& 0 \, .
\end{eqnarray}
From here we can compute the velocity as defined in Eq.~\eqref{contra-vel},
\begin{eqnarray} \label{phase-vel-PX}
    {\bf v}^2 = 1 - \frac{2 c_2 (\partial \phi \cdot N)^2}{(1 - 3 c_2 (\partial \phi)^2) (N^0)^2} \, ,
\end{eqnarray}
which coincides with the usual definition of phase velocity $\omega^2({\bf p})/{\bf p}^2$ for a background with a spatially isotropic acoustic metric.\footnote{See the discussion in \cite{Sawicki:2024ryt}. If we impose $N^\mu p_\mu = 0$, then it follows that $\frac{{\bf N} \cdot {\bf p}}{N^0 \omega} = 1$, meaning that $\omega^2/{\bf p}^2={\bf N}^2/(N^0)^2$, provided ${\bf p}$ is parallel to ${\bf N}$.}

Suppose that in this theory we want to avoid superluminality. Requiring the velocity \eqref{phase-vel-PX} to be less or equal to the speed of light $c=1$ immediately implies\footnote{Here we are assuming that the denominator above is always positive, which is guaranteed by the absence-of-ghosts condition.} 
\begin{eqnarray}\label{c2-bound-PX}
    c_2 \geq 0 \, .
\end{eqnarray}
This is the well known condition to prevent superluminality in this model \cite{Adams:2006sv}. Geometrically, this means the acoustic vector $N^\mu$ must be timelike (or null) according to the standard spacetime metric $\eta_{\mu\nu}$, i.e. lie inside (or on) the lightcone. Using Eq.~\eqref{N-EoM-PX} above
\begin{eqnarray}
    \eta_{\mu\nu} N^\mu N^\nu = - \frac{2 c_2 \, (\partial{\phi} \cdot N)^2}{\left ( 1 - 3 c_2 (\partial  {\phi})^2 \right)} \leq 0 \, ,
\end{eqnarray}
which again implies \eqref{c2-bound-PX}. 

As a closing remark, in this case due to the absence of a mass term, we could have also obtained this result by instead demanding that the momentum co-vector $p_\mu$ lies outside (or on) the lightcone, that is, it is $\eta$-spacelike (or null). From the EoM in terms of $p_\mu$, we get
\begin{eqnarray}
    \eta^{\mu\nu} p_\mu p_\nu = \frac{2 c_2 \, (\partial{\phi} \cdot p)^2}{\left ( 1 - c_2 (\partial  {\phi})^2 \right)} \geq 0 \, ,
\end{eqnarray}
again leading to \eqref{c2-bound-PX}. However, in general this does not imply that the velocity in Eq.~\eqref{contra-vel} agrees with $\omega^2({\bf p})/{\bf p}^2$. This is especially important when comparing it against $c_s^2 \neq 1$ or when handling a theory with a mass gap.


\bibliographystyle{JHEP}
\bibliography{biblio.bib}

\end{document}